\documentclass[%
 reprint,
 amsmath,amssymb,
 aps,
showkeys
]{revtex4-1}

\usepackage{graphicx}
\usepackage{dcolumn}
\usepackage{bm}

\begin{document}


\title{Parameterizing the Energy Dissipation Rate in Stably Stratified Flows}

\author{Sukanta Basu}
\email{sukanta.basu@gmail.com}
\affiliation{Faculty of Civil Engineering and Geosciences, Delft University of Technology, Delft, the Netherlands
}
\author{Ping He}%
\email{drpinghe@umich.edu}
\affiliation{Department of Aerospace Engineering, University of Michigan, Ann Arbor, USA}

\author{Adam W. DeMarco}
\email{awdemarc@ncsu.edu}
\affiliation{United States Air Force, USA }%
\date{\today}

\begin{abstract}
We use a database of direct numerical simulations to evaluate parametrizations for energy dissipation rate in stably stratified flows. We show that shear-based formulations are more appropriate for stable boundary layers than commonly used buoyancy-based formulations. As part of the derivations, we explore several length scales of turbulence and investigate their dependence on local stability.  
\end{abstract}

\maketitle


\section{\label{sec:level1}Introduction}

Energy dissipation rate is a key variable for characterizing turbulence \citep{vassilicos15}. It is a sink term in the prognostic equation of turbulence kinetic energy (TKE; $\overline{e}$): 
\begin{equation}
    \frac{\partial \overline{e}}{\partial t} + ADV = BNC + SHR + TRP + PRC - \overline{\varepsilon},
    \label{TKE}
\end{equation}
where, $\overline{\varepsilon}$ is the mean energy dissipation rate. The terms $ADV,$ $BNC,$ $SHR,$ $TRP,$ and $PRC$ refer to advection, buoyancy production (or destruction), shear production, transport, and pressure correlation terms, respectively. Energy dissipation rate also appears in the celebrated ``-5/3 law'' of \citeauthor{kolmogorov41a}~\cite{kolmogorov41a} and \citeauthor{obukhov41a}~\cite{obukhov41a,obukhov41b}: 
\begin{equation}
    E(\kappa) \approx \overline{\varepsilon}^{2/3} \kappa^{-5/3},
    \label{K41}
\end{equation}
where, $E(\kappa)$ and $\kappa$ denote the energy spectrum and wavenumber, respectively. 

In field campaigns or laboratory experiments, direct estimation of $\overline{\varepsilon}$ has always been a challenging task as it involves measurements of nine components of the strain rate tensor. Thus, several approximations (e.g., isotropy, Taylor's hypothesis) have been utilized and a number of indirect measurement techniques (e.g., scintillometers, lidars) have been developed over the years. In parallel, a significant effort has been made to correlate $\overline{\varepsilon}$ with easily measurable meteorological variables. For example, several flux-based and gradient-based similarity hypotheses have been proposed \citep[e.g.,][]{wyngaard71a,wyngaard71b,thiermann92,hartogensis05}. 

In addition, a handful of papers also attempted to establish relationships between $\overline{\varepsilon}$ and either the vertical velocity variance ($\sigma_w^2$) or TKE ($\overline{e}$). One of the first relationships was proposed by \citeauthor{chen74}~\cite{chen74}. By utilizing the Kolmogorov\textendash Obukhov spectrum (i.e., Eq.~\ref{K41}) with certain assumptions, he derived: 
\begin{equation}
    \overline{\varepsilon} \propto \sigma_w^3, 
    \label{C74}
\end{equation}
where, the proportionality constant is not dimensionless. Since this derivation is only valid in the inertial range of turbulence, a band-pass filtering of vertical velocity measurements was recommended prior to computing $\sigma_w$. A few years later, \citeauthor{weinstock81}~\cite{weinstock81} revisited the work of \cite{chen74} and again made use of Eq.~\ref{K41}, albeit with different assumptions (see Appendix~2 for details). He arrived at the following equation:
\begin{equation}
    \overline{\varepsilon} \approx \sigma_w^2 N, 
    \label{W81}
\end{equation}
where, $N$ is the so-called Brunt\textendash V\"{a}is\"{a}la frequency. Using observational data from the stratosphere, \citeauthor{weinstock81}~\cite{weinstock81} demonstrated the superiority of Eq.~\ref{W81} over Eq.~\ref{C74}. In a recent empirical study, by analyzing measurements from the CASES-99 (the Cooperative Atmosphere\textendash Surface Exchange Study\textendash 1999) field campaign, \citeauthor{bocquet11}~\cite{bocquet11} proposed to use $\overline{\varepsilon}$ as a proxy for $\sigma_w^2$. 

In the present work, we quantify the relationship between $\overline{\varepsilon}$ and $\overline{e}$ (as well as between $\overline{\varepsilon}$ and $\sigma_w$) by using turbulence data generated by direct numerical simulation (DNS). To this end, we first compute several well-known ``outer'' length scales (e.g., buoyancy length scale and Ozmidov scale), normalize them appropriately, and explore their dependence on height-dependent stability. Next, we investigate the inter-relationships of certain (normalized) outer length scales (OLS) that portray qualitatively similar stability-dependence. By analytically expanding these relationships, we arrive at two $\overline{\varepsilon}$--$\overline{e}$ and two  $\overline{\varepsilon}$--$\sigma_w$ formulations; only the shear-based formulations portray quasi-universal scaling. 

The organization of this paper is as follows. In Sect.~2, we describe our DNS runs and subsequent data analyses. Simulated results pertaining to various length scales are included in Sect.~3. The $\overline{\varepsilon}$--$\overline{e}$ and $\overline{\varepsilon}$--$\sigma_w$ formulations are derived in Sect.~4. We discuss the surface-layer characteristics of a specific shear-based length scale in Sect.~5. A few concluding remarks, including the implications of our results for atmospheric modelling, are made in Sect.~6. In order to enhance the readability of the paper, either a heuristic or an analytical derivation of all the length scales is provided in Appendix~1. Given the importance of Eq.~\ref{W81}, its derivation is also summarized in Appendix~2. In Appendix~3, we elaborate on the normalization of various variables that are essential for the post-processing of DNS-generated data. Finally, supplementary results based on our DNS database are included in Appendix~4.  

\section{Direct Numerical Simulation}

Over the past decade, due to the increasing abundance of high-performance computing resources, several studies probed different types of stratified flows by using DNS \citep[e.g.,][]{flores11,garcia11,brethouwer12,chung12,ansorge14,shah14,he15,he16b}. These studies provided valuable insights into the dynamical and statistical properties of these flows (e.g., intermittency, structure parameters). In the present study, we use a DNS database, which was previously generated by using a massively parallel DNS code, called HERCULES \citep{he16a}, for the parametrization of optical turbulence \citep{he16c}. 
The verification of HERCULES has been conducted in the appendix of ~\cite{he16a}. 
We solved the normalized Navier\textendash Stokes and temperature equations in an open channel driven by a streamwise pressure gradient, as shown in Eqs.~\ref{eqn_mass}\textendash \ref{eqn_energy} (using Einstein's summation notation for subscripts $i$ and $j$): 

\begin{widetext}
\begin{equation}
\label{eqn_mass}
\frac{\partial u_{n, i}}{\partial x_{n, i}} =0,
\end{equation}

\begin{equation}
\label{eqn_momentum}
\frac{\partial u_{n, i}}{\partial t_n}+\frac{\partial u_{n, i} u_{n, j}}{\partial x_{n, j}}=-\frac{\partial p_n}{\partial x_{n, i}} +\frac{1}{Re_b} \frac{\partial }{\partial x_{n, j}} \left(\frac{\partial u_{n, i}}{\partial x_{n, j}}\right) + \mathrm{\Delta}P {\delta}_{i1}+Ri_b {\theta_n} {\delta}_{i3}, 
\end{equation}

\begin{equation}
\label{eqn_energy}
\frac{\partial \theta_n}{\partial t_n}+\frac{\partial \theta_n u_{n, i}}{\partial x_{n, i}}=\frac{1}{Re_b Pr} \frac{\partial }{\partial x_{n, i}} \left(\frac{\partial \theta_n}{\partial x_{n, i}}\right),
\end{equation}
\end{widetext}
where $u_n$ and $x_n$ are the normalized velocity and coordinate vectors, respectively, with the subscript $i$ denoting the $i$\textsuperscript{th} vector component; $t_n$ is the normalized time; $p_n$ is the normalized pressure; $\mathrm{\Delta}P$ is the streamwise pressure gradient to drive the flow; and $\theta_n$ is the normalized potential temperature. The normalization of DNS variables is shown in Appendix 3. Throughout the paper, the subscript ``$n$'' is used to denote a normalized variable.

The computational domain size for all the DNS runs was $L_x \times L_y \times L_z = 18 h \times 10 h \times h$, where $h$ is the open-channel height. The domain was discretized by $2304 \times 2048 \times 288$ grid points in streamwise, spanwise, and wall-normal directions, respectively. The bulk Reynolds number, $Re_b$, for all the simulations was fixed at 20000, defined as: 
\begin{equation}
Re_b = U_b h/\nu,
\end{equation}
where, $\nu$ and $U_b$ denote kinematic viscosity and the bulk (averaged) velocity in the channel, respectively. The constant $Re_b$ was achieved by dynamically adjusting $\mathrm{\Delta} P$ in Eq.~\ref{eqn_mass} during the simulations. The corresponding friction Reynolds number ($Re_\tau$) ranges from 575 to 902. The bulk Richardson number was calculated as: 
\begin{equation}
Ri_b = \frac{\left(\theta_{top}-\theta_{bot}\right)g h}{U_b^2 \theta_{top}},
\end{equation}
where $\theta_{top}$ and $\theta_{bot}$ represent potential temperature at the top and the bottom of the channel, respectively. The gravitational acceleration is denoted by $g$. 

A total of five simulations were performed with gradual increase in the temperature difference between the top and bottom walls (effectively by increasing $Ri_b$) to mimic the night-time cooling of the land-surface. The normalized cooling rates ($CR$), $\partial Ri_b/\partial T_n$, ranged from $1\times10^{-3}$ to $5\times10^{-3}$, where $T_n$ is a non-dimensional time ($=tU_b/h$). All our simulations started with neutral conditions: $Ri_b = 0$. After $T_n = 100$, each simulation evolved to a different $Ri_b$ value, ranging from 0.1 to 0.5. Since we were considering atmospheric flows, the Prandtl number, $Pr = \nu/k$ was assumed to be equal to 0.7, with $k$ being the thermal diffusivity.  

The simulation results were output every 10 non-dimensional time. To avoid spin-up issues, in the present study, we only use data for the last five output files (i.e., $60 \le T_n \le 100$). Furthermore, we only consider data from the region $0.1 h\le z \le 0.5 h$ to discard any blocking effect of the surface or avoid any laminarization in the upper part of the open channel. 

The turbulence kinetic energy and its mean dissipation are computed as follows (using Einstein's summation notation): 
\begin{subequations}
\begin{equation}
\overline{e} = \frac{1}{2} \overline{u_i' u_i'},
\end{equation}
\begin{equation}
\overline{\varepsilon} = \nu \left(\overline{\frac{\partial u_i'}{\partial x_j} \frac{\partial u_i'}{\partial x_j}}\right).
\end{equation}
\end{subequations}
In these equations and in the rest of the paper, the ``overbar'' notation is used to denote mean quantities. Horizontal (planar) averaging operation is performed for all the cases. The ``prime'' symbol is used to represent the fluctuation of a variable with respect to its planar averaged value. 

\begin{figure*}[htbp]
\centering
  \includegraphics[width=2.3in,height=2.3in]{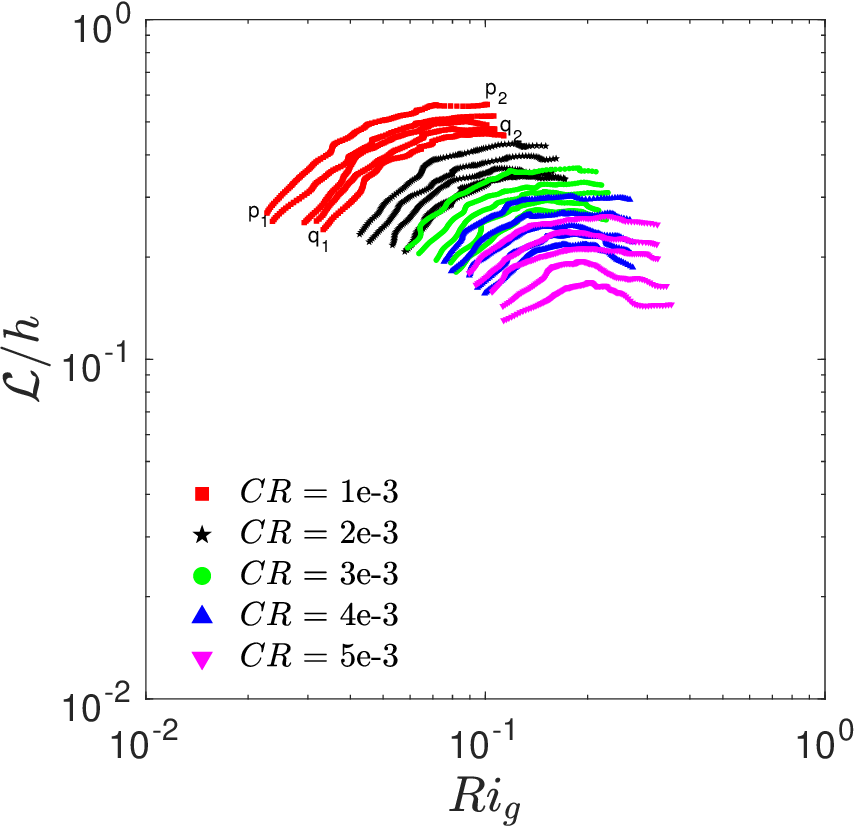}
  \hspace{0.3in}
  \includegraphics[width=2.3in,height=2.3in]{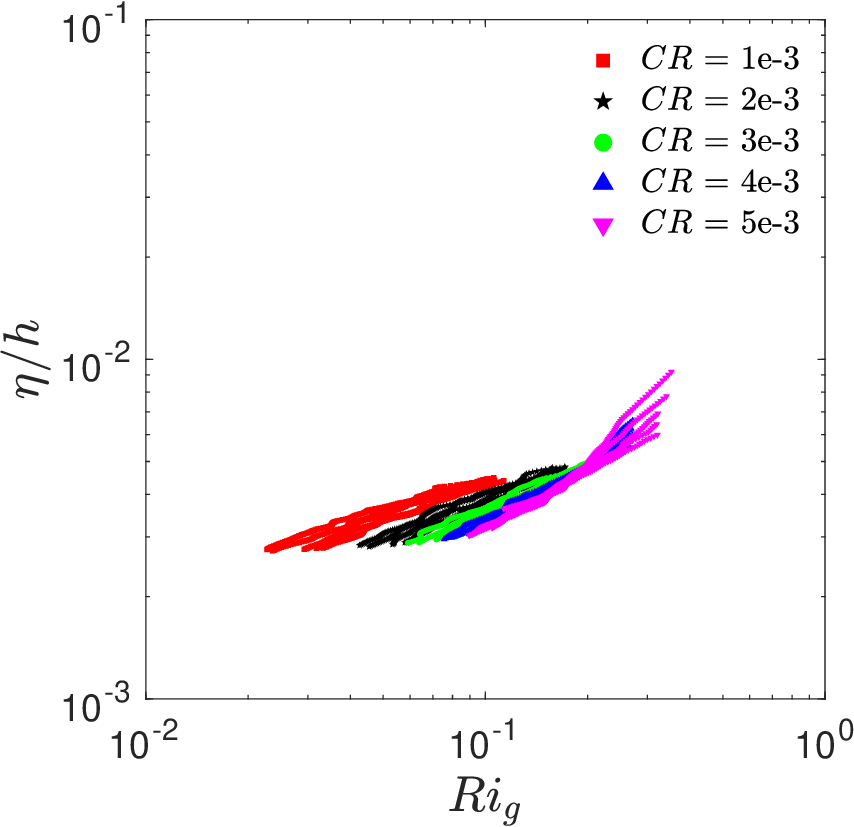}
\caption{Integral (left panel) and Kolmogorov (right panel) length scales as functions of gradient Richardson number. Both the length scales are normalized by the height of the open channel ($h$). Simulated data from five different DNS runs are represented by different coloured symbols in these plots. In the legends, $CR$ represents normalized cooling rates. The points $p_1$ and $p_2$ represent data from $z/h = 0.1$ and $z/h = 0.5$, respectively, at non-dimensional time ($T_n$) of 60. Similarly, $q_1$ and $q_2$ are associated with data from $z/h = 0.1$ and $z/h = 0.5$, respectively, at non-dimensional time ($T_n$) of 100}
\label{fig1}      
\end{figure*}

\section{Length Scales}

In this section, we discuss various length scales of turbulence. To enhance the readability of the paper, we do not elaborate on their derivations or physical interpretations here; for such details, the readers are directed to Appendix~1. 

From the DNS-generated data, we first calculate the integral length scale ($\mathcal{L}$) and Kolmogorov length scale ($\eta$). They are defined as \citep{tennekes72,pope00}: 
\begin{subequations}
\begin{equation}
    \mathcal{L} \equiv \frac{\overline{e}^{3/2}}{\overline{\varepsilon}},
\end{equation}
\begin{equation}
    \eta \equiv \left(\frac{\nu^3}{\overline{\varepsilon}}\right)^{1/4}.
\end{equation}
\end{subequations}
In Fig.~\ref{fig1}, normalized values of $\mathcal{L}$ and $\eta$ are plotted against the gradient Richardson number ($Ri_g = N^2/S^2$), where $S$ is the magnitude of wind shear. $N$ and $S$ are computed as follows: 
\begin{subequations}
\begin{equation}
    N = \sqrt{ \frac{g}{\Theta_0} \frac{\partial \overline{\theta}}{\partial z} },
\end{equation}
\begin{equation}
    S = \sqrt{ \left(\frac{\partial \overline{u}}{\partial z}\right)^2 + \left(\frac{\partial \overline{v}}{\partial z}\right)^2 },
\end{equation}
\end{subequations}
where, $\Theta_0$ is a reference temperature. As mentioned earlier, the overbar denotes horizonal (planar) averaging operation. In the  left panel, we marked four specific points based on the data from DNS run with imposed cooling rate of 10$^{-3}$ to better understand the effects of height and stability on the integral length scale. The points $p_1$ and $p_2$ represent data from $z/h = 0.1$ and $z/h = 0.5$, respectively, at non-dimensional time ($T_n$) of 60. Similarly, $q_1$ and $q_2$ are associated with data from $z/h = 0.1$ and $z/h = 0.5$, respectively, at non-dimensional time ($T_n$) of 100.

Physically, one would expect the integral scale to increase with height as long as the eddies feel the presence of the surface (near-neutral or weakly stable condition). For very stable conditions, the eddies no longer feel the presence of the surface. In the atmospheric boundary layer literature, it is known as the z-less condition \citep{wyngaard73,grisogono10}. Under the influence of strong stability, the integral length scales become more or less independent of the height above the surface.

From Fig.~\ref{fig1}, it is clear that the integral length scale increases with height and slowly decreases with time in all the simulations due to the increasing stability effects. Simulations with higher cooling rates have smaller integral length scales. Some of these runs (e.g., $CR = 5\times10^{-3}$) exhibit z-less behaviour due to strong stability effects.

In contrast, $\eta$ marginally increases with higher stability due to lower $\overline{\varepsilon}$. The ratio of $\mathcal{L}$ to $\eta$ decreases from about 100 to 20 as stability is increased from a weakly stable condition to a strongly stable condition.

Next, we compute four outer length scales: Ozmidov ($L_{OZ}$), Corrsin ($L_C$), buoyancy ($L_b$), and Hunt ($L_H$). They are defined as \citep{corrsin58,dougherty61,ozmidov65,brost78,hunt88,hunt89,sorbjan08,wyngaard10}: 
\begin{subequations}
\begin{equation}
    L_{OZ} \equiv \left(\frac{\overline{\varepsilon}}{N^3}\right)^{1/2},
    \label{OLSa}
\end{equation}
\begin{equation}
    L_C \equiv \left(\frac{\overline{\varepsilon}}{S^3}\right)^{1/2},
    \label{LC}
\end{equation}
\begin{equation}
    L_b \equiv \frac{\overline{e}^{1/2}}{N},
\end{equation}
\begin{equation}
    L_H \equiv \frac{\overline{e}^{1/2}}{S}.
    \label{OLSd}
\end{equation}
\label{OLS}
\end{subequations}

\noindent Please note that, in the literature, $L_b$ and $L_H$ have also been defined as $\sigma_w/N$ and $\sigma_w/S$, respectively. Both $L_{OZ}$ and $L_C$ are functions of $\overline{\varepsilon}$, a microscale variable. In contrast, $L_b$ and $L_H$ only depend on macroscale variables. 

Both shear and buoyancy prefer to deform the larger eddies compared to the smaller ones \citep{itsweire93,smyth00,chung12,mater13}. Eddies that are smaller than $L_C$ or $L_H$ are not affected by shear. Similarly, buoyancy does not influence the eddies of size less than $L_{OZ}$ or $L_b$. In other words, the eddies can be assumed to be isotropic if they are smaller than all these OLSs.  

\begin{figure*}[ht]
\centering
  \includegraphics[width=2.3in,height=2.3in]{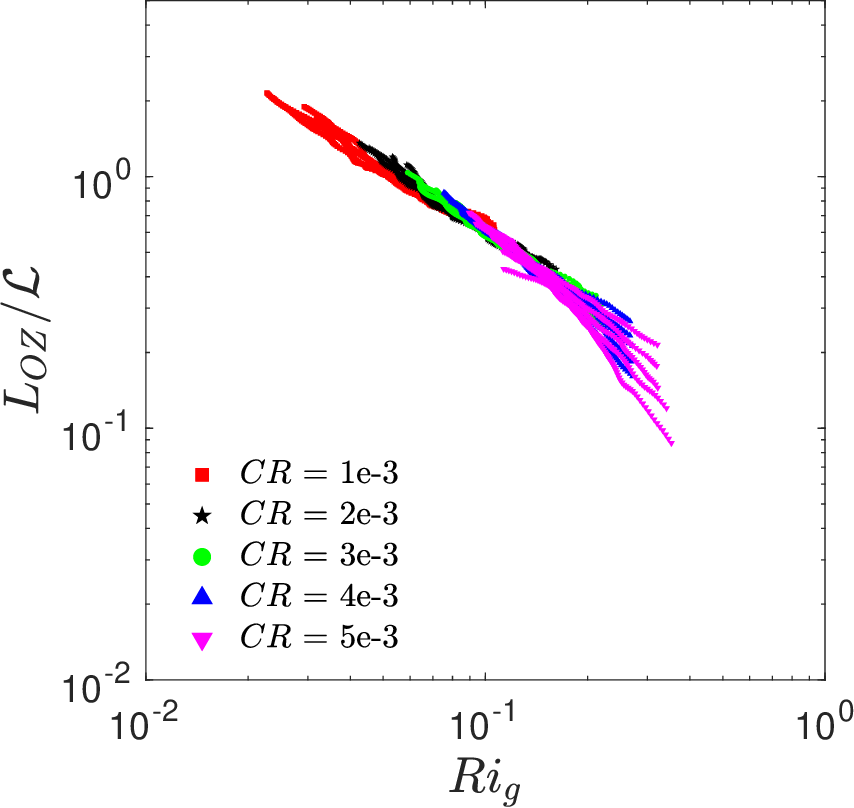}
  \hspace{0.3in}
  \includegraphics[width=2.3in,height=2.3in]{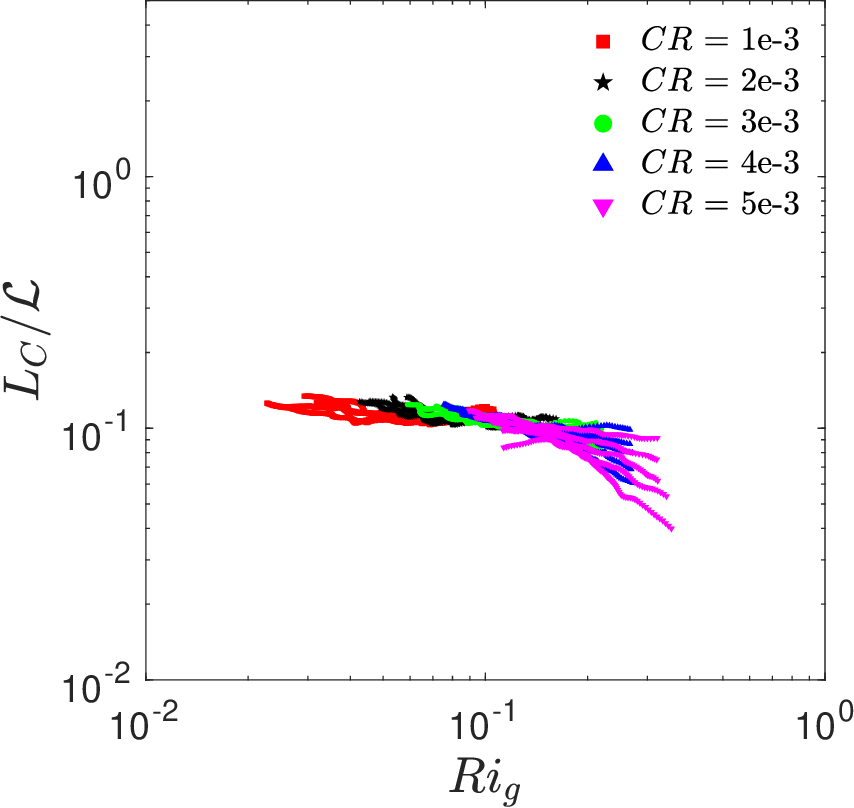}\\
  \includegraphics[width=2.3in,height=2.3in]{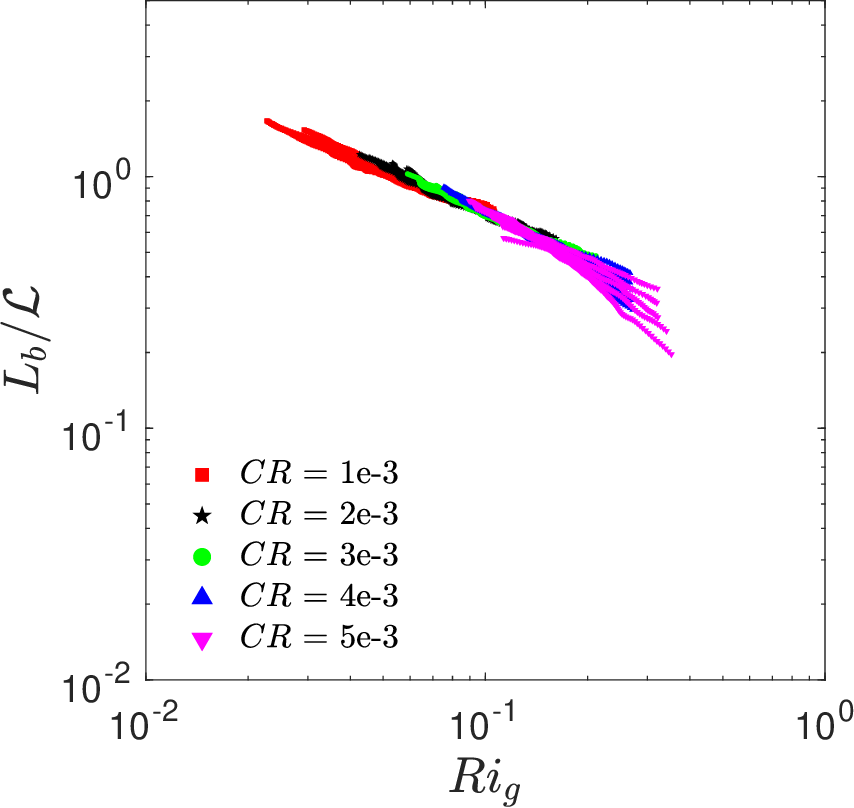}
  \hspace{0.3in}
  \includegraphics[width=2.3in,height=2.3in]{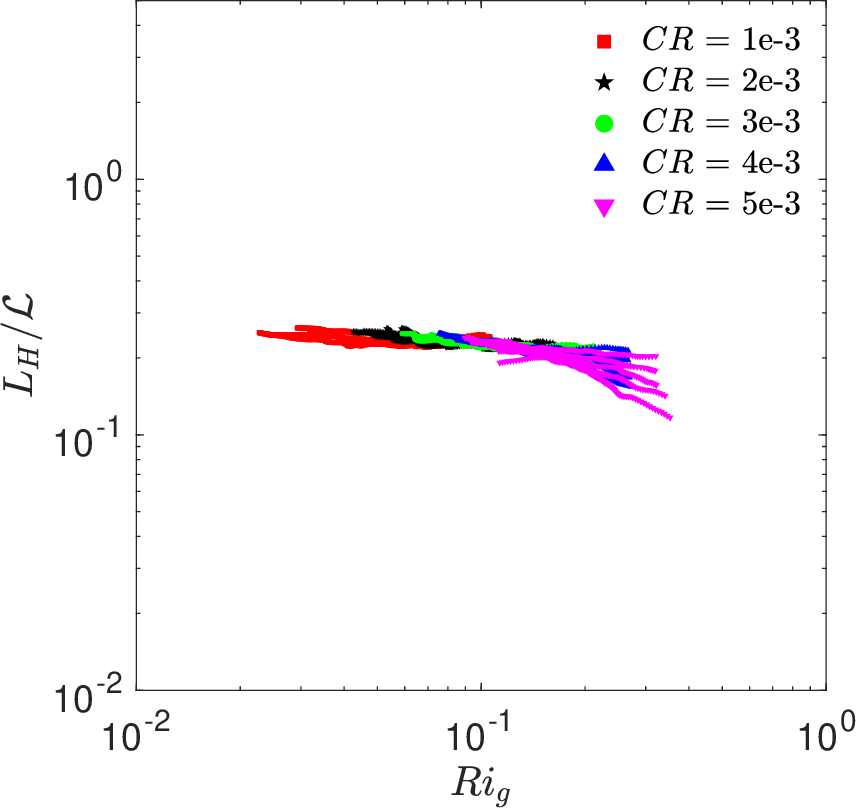}
\caption{Ozmidov (top-left panel), Corrsin (top-right panel), buoyancy (bottom-left panel), and Hunt (bottom-right panel) length scales as functions of gradient Richardson numbers. All the length scales are normalized by the integral length scale. Simulated data from five different DNS runs are represented by different coloured symbols in these plots. In the legends, $CR$ represents normalized cooling rates}
\label{fig2}      
\end{figure*}

Since $\mathcal{L}$ changes across the simulations, all the OLS values are normalized by corresponding $\mathcal{L}$ values and plotted as functions of $Ri_g$ in Fig.~\ref{fig2}. The collapse of the data from different runs, on to seemingly universal curves, is remarkable for all the cases except for $Ri_g > 0.2$. We would like to mention that similar scaling behaviour was not found if other normalization factors were used. For instance, we have tried the height of the open channel ($h$) as a normalization factor. We also tested several definitions of the boundary-layer height (e.g., the height where variances or fluxes decrease to a small percentage of the peak magnitude). None of them resulted in any scaling relationship.     

Both normalized $L_{OZ}$ and $L_b$ decrease monotonically with $Ri_g$; however, the slopes are quite different. The length scales $L_C$ and $L_H$ barely exhibit any sensitivity to $Ri_g$ (except for $Ri_g > 0.1$). Even for weakly stable conditions, these length scales are less than 25 percent of $\mathcal{L}$. 

Based on the expressions of the OLSs (i.e., Eq.~\ref{OLS}a-d) and the definition of the gradient Richardson number, we can write: 
\begin{subequations}
\begin{equation}
    \frac{L_C}{L_{OZ}} = \left(\frac{N}{S}\right)^{3/2} = Ri_g^{3/4},
\end{equation}
\begin{equation}
    \frac{L_H}{L_{b}} = \left(\frac{N}{S}\right) = Ri_g^{1/2}. 
\end{equation}
\end{subequations}
Thus, for $Ri_g < 1$, one expects $L_C < L_{OZ}$ and $L_H < L_b$. Such relationships are fully supported by Fig.~\ref{fig2}. In comparison to the buoyancy effects, the shear effects are felt at smaller length scales for the entire stability range considered in the present study. 

\begin{figure*}[ht]
\centering
  \includegraphics[width=2.3in,height=2.3in]{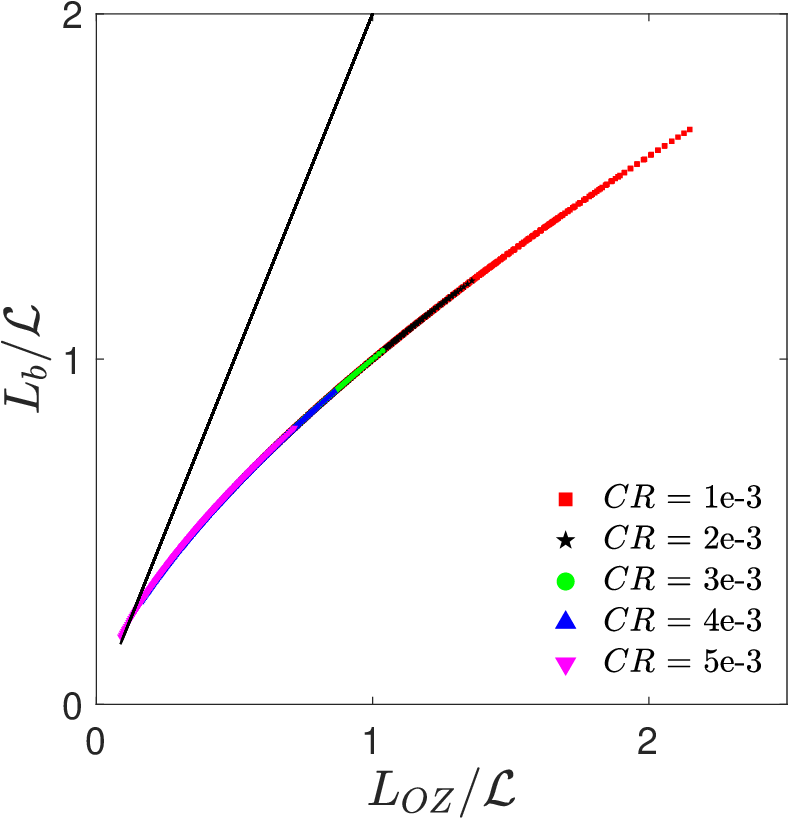}
  \hspace{0.3in}
  \includegraphics[width=2.3in,height=2.3in]{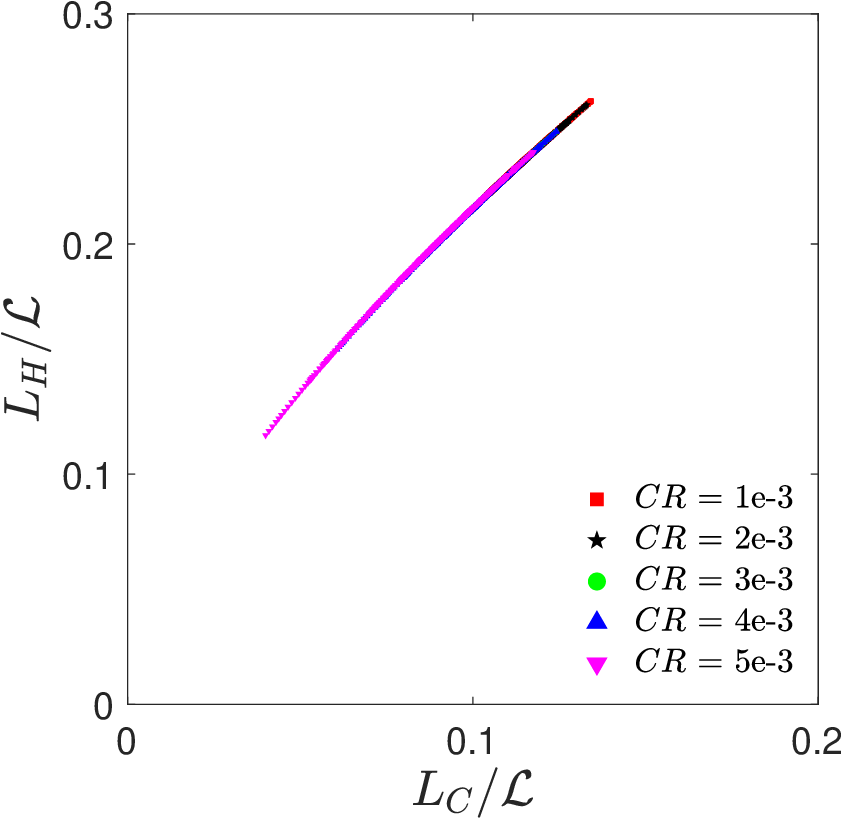}
\caption{Left panel: variation of the normalized buoyancy length scale against the normalized Ozmidov length scale. Right panel: variation of the normalized Hunt length scale against the normalized Corrsin length scale. Simulated data from five different DNS runs are represented by different coloured symbols in these plots. In the legends, $CR$ represents normalized cooling rates}
\label{fig3}       
\end{figure*}

Owing to their similar scaling behaviours, $L_b/\mathcal{L}$ against $L_{OZ}/\mathcal{L}$ are plotted in Fig.~\ref{fig3} (left panel). Once again, all the simulated data collapse nicely in a quasi-universal (nonlinear) curve. Since in a double-logarithmic representation (not shown) this curve is linear, we can write: 
\begin{equation}
    \frac{L_b}{\mathcal{L}} \equiv \left(\frac{L_{OZ}}{\mathcal{L}}\right)^m,
    \label{Lb_vs_LOZ}
\end{equation}
where, $m$ is an unknown power-law exponent. Via regression analysis, we estimate $m = 2/3$. By using $L_b \equiv \overline{e}^{1/2}/N$ and the definitions of $L_{OZ}$ and $\mathcal{L}$, we arrive at: 
\begin{equation}
    \frac{\overline{e}^{1/2}}{N} = \left(\frac{\overline{\varepsilon}}{N^3} \right)^{m/2} \left(\frac{\overline{e}^{3/2}}{\overline{\varepsilon}} \right)^{1-m}.
\end{equation}
Further simplification leads to: $\overline{\varepsilon} = \overline{e} N$; please note that the exponent $m$ cancels out in the process. Instead of $\overline{e}^{1/2}$, if we utilize $\sigma_w$ in the definitions of $L_b$ and $\mathcal{L}$, we get: $\overline{\varepsilon} = \sigma_w^2 N$. This equation is identical to Eq.~\ref{W81}, which was derived by \citeauthor{weinstock81}~\cite{weinstock81}. His derivation, based on inertial-range scaling, is summarized in Appendix~2.

In the right panel of Fig.~\ref{fig3}, we plot $L_H/\mathcal{L}$ versus $L_{C}/\mathcal{L}$. Both these normalized length scales have limited ranges; nonetheless, they are proportional to one another. Like Eq.~\ref{Lb_vs_LOZ}, we can write in this case: 
\begin{equation}
    \frac{L_H}{\mathcal{L}} \equiv \left(\frac{L_{C}}{\mathcal{L}}\right)^n,
    \label{LH_vs_LC}
\end{equation}
where, $n$ estimated via regression analysis is also found to be equal to 2/3. The expansion of this equation leads to either $\overline{\varepsilon} = \overline{e} S$ or $\overline{\varepsilon} = \sigma_w^2 S$, depending on the definition of $L_H$ and $\mathcal{L}$.

\section{Parameterizing the Energy Dissipation Rate}

\begin{figure*}
\centering
  \includegraphics[width=2.3in,height=2.3in]{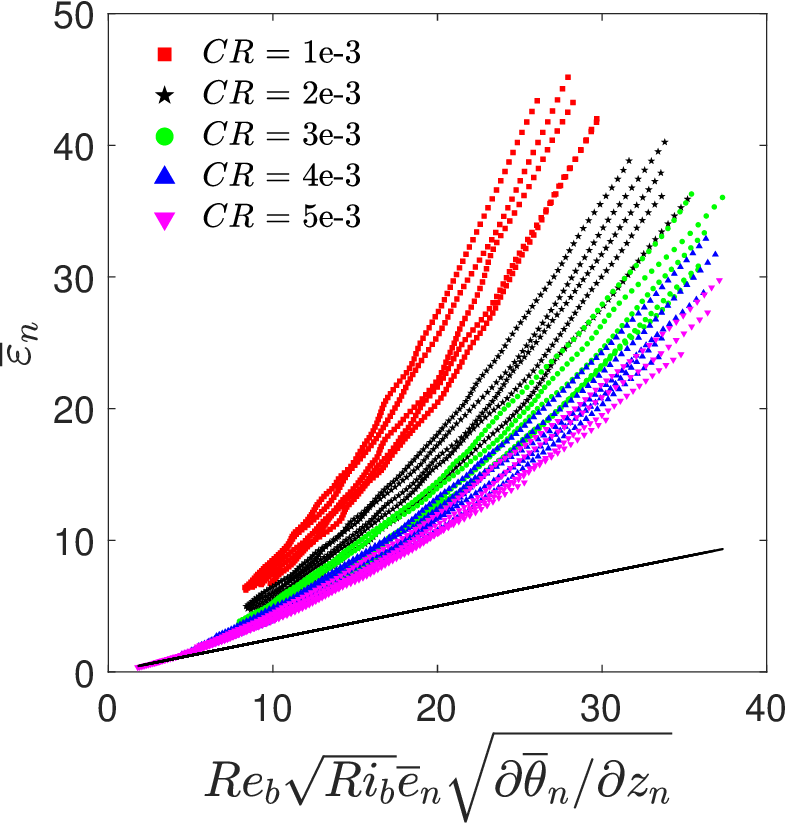}
  \hspace{0.3in}
  \includegraphics[width=2.3in,height=2.3in]{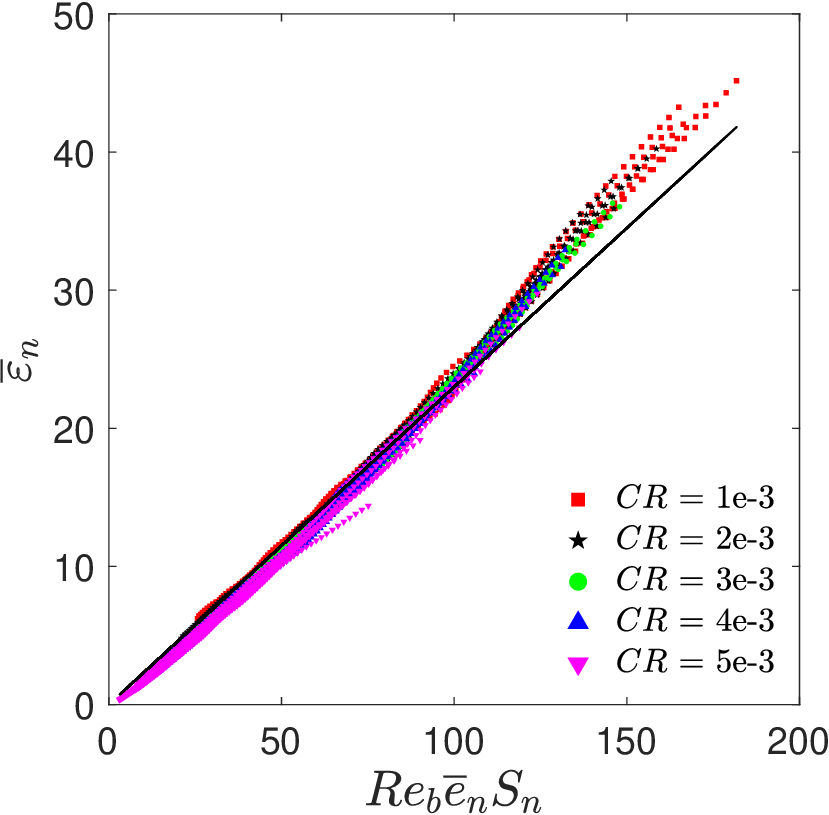}\\
  \includegraphics[width=2.3in,height=2.3in]{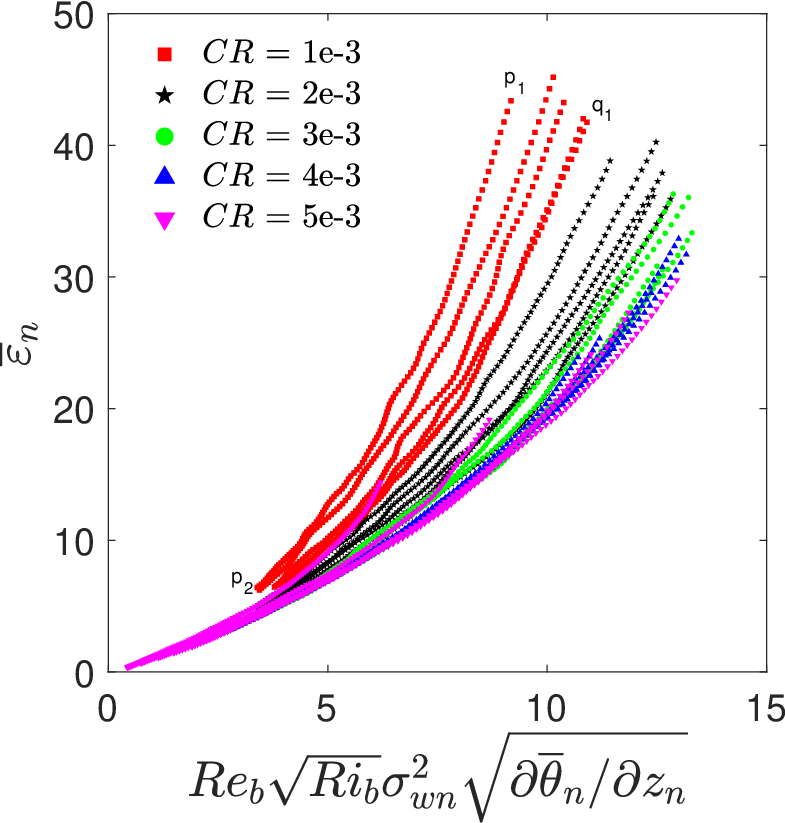}
  \hspace{0.3in}
  \includegraphics[width=2.3in,height=2.3in]{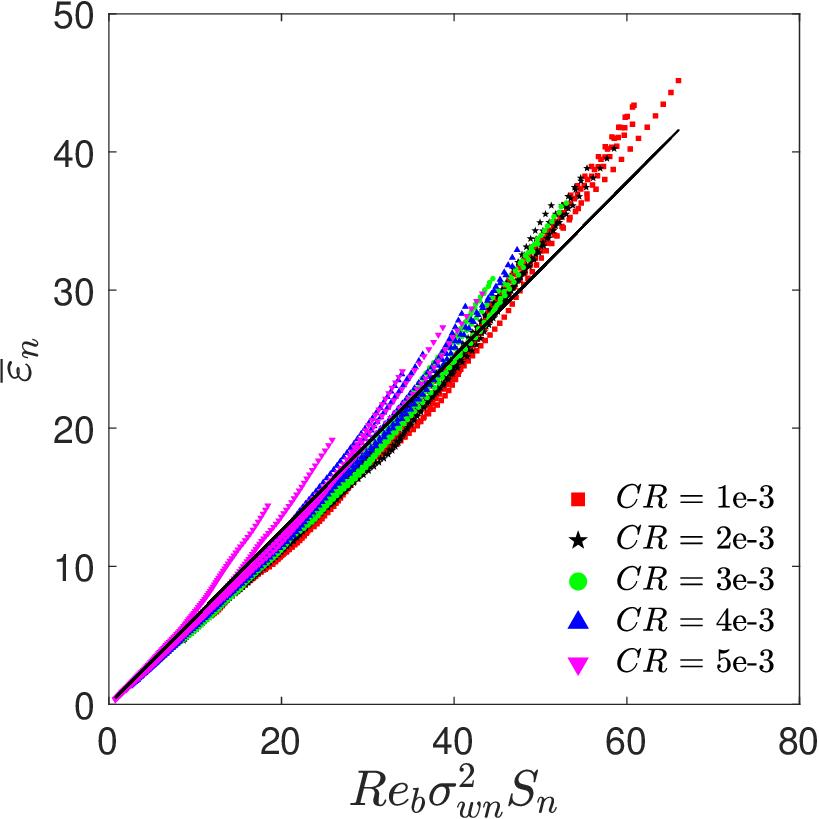}
\caption{Variation of normalized energy dissipation rates against normalized $\overline{e} N$ (top-left panel), normalized $\overline{e} S$ (top-right panel), normalized $\sigma_w^2 N$ (bottom-left panel), and normalized $\sigma_w^2 S$ (bottom-right panel). Simulated data from five different DNS runs are represented by different coloured symbols in these plots. In the legends, $CR$ represents normalized cooling rates. In the bottom-left panel, the points $p_1$ and $p_2$ represent data from $z/h = 0.1$ and $z/h = 0.5$, respectively at non-dimensional time ($T_n$) of 60. Whereas, $q_1$ is associated with data from $z/h = 0.1$ at non-dimensional time ($T_n$) of 100}
\label{fig4}       
\end{figure*}

Earlier in Fig.~\ref{fig3}, we plotted normalized OLS values against one another. It is plausible that the apparent data collapse is simply due to self-correlation as the same variables (i.e., $\mathcal{L}$, $N$, and $S$) appear in both abscissa and ordinate. To further probe into this problematic issue, we produce Fig.~\ref{fig4}. Here, we basically plot normalized $\overline{\varepsilon}$ as functions of normalized $\overline{e} N$, $\overline{e} S$, $\sigma_w^2 N$, and $\sigma_w^2 S$, respectively. These plots have completely independent abscissa and ordinate terms and do not suffer from self-correlation. Please note that the appearance of $Re_b$ and $Ri_b$ in these figures is due to the normalization of variables in DNS. The definitions of all the normalized variables (e.g., $\overline{\varepsilon}_n$) are provided in Appendix~3. 

It is clear that the plots in the left panel of Fig.~\ref{fig4}, which involve $N$, do not show any universal scaling. For low $CR$ values, normalized $\overline{\varepsilon}$ values do not go to zero; this behaviour is physically realistic. One cannot expect $\overline{\varepsilon}$ to go to zero for neutral condition (i.e., $N \to 0$). With increasing cooling rates, the curves seem to converge to an asymptotic curve that passes through the origin. As $\overline{e}$ or $\sigma_w$ continually reduces with increasing stability, one does expect $\overline{\varepsilon}$ to approach zero. 

In a seminal paper, \citeauthor{deardorff80}~\cite{deardorff80} proposed a parametrization for $\overline{\varepsilon}$, which for strongly stratified conditions approaches $0.25 \overline{e} N$. In Fig.~\ref{fig4} (top-left panel), we overlaid $\overline{\varepsilon} = 0.25 \overline{e} N$ on the DNS-generated data. Clearly, it only overlaps with the simulated data at the strongly stratified region. If $\overline{\varepsilon} = 0.25 \overline{e} N$ is used in the definition of $L_{OZ}$, after simplification, one gets $L_{OZ} = L_b/2$. The line $L_b = 2 L_{OZ}$ is drawn in Fig.~\ref{fig3}. As would be anticipated, it only overlaps with the simulated data when the OLS values are the smallest (signifying strongly stable condition).     

Compared to the left panels, the right panels of Fig.~\ref{fig4} portray very different scaling characteristics. All the data collapse on quasi-universal curves remarkably, especially, for the $\overline{\varepsilon} \approx \overline{e} S$ case. The slopes of the regression lines, estimated via conventional least-squares approach and bootstrapping \citep{efron82,mooney93}, are shown on these plots. Essentially, we have found:
\begin{subequations}
\begin{equation}
    \overline{\varepsilon} = 0.23 \overline{e} S,
    \label{EDR1}
\end{equation}

\begin{equation}
    \overline{\varepsilon} = 0.63 \sigma_w^2 S.
    \label{EDR2}
\end{equation}
\end{subequations}
We note that our estimated coefficient 0.63 is within the range of values reported by \cite{schumann95} from laboratory experiments and large-eddy simulations (please refer to their Fig.~1). 

In summary, neither $\overline{\varepsilon} = \overline{e} N$ nor $\overline{\varepsilon} = \sigma_w^2 N$ are appropriate parametrizations for weakly or moderately stratified conditions; they may provide reasonable predictions for very stable conditions. In contrast, the shear-based parametrizations should be applicable from a wide range of stability conditions, from near-neutral to at least $Ri_g \approx 0.2$. Since within the continuously turbulent stable boundary layer (SBL), $Ri_g$ rarely exceeds 0.2 \citep[see][]{garratt82,nieuwstadt84}, we believe Eq.~\ref{EDR1} or Eq.~\ref{EDR2} will suffice for most practical boundary-layer applications. However, for intermittently turbulent SBLs and the free atmosphere, where $Ri_g$ can exceed O(1), Deardorff's parametrization (i.e., $\overline{\varepsilon} = 0.25 \overline{e} N$) might be a more viable option. Unfortunately, we cannot verify this speculation using our existing DNS dataset.   

\section{Discussions}

\citeauthor{hunt88}~\cite{hunt88,hunt89} stated that $L_H$ may not be a representative length scale near the surface due to the blocking effect. From our perspective, $L_H$ does possess the correct surface-layer characteristics, as elaborated below.

Following Nieuwstadt's local scaling \citep{nieuwstadt84} and Monin\textendash Obukhov similarity theory, we can rewrite $L_H$ as follows for the surface layer:
\begin{equation}
    L_H \equiv \frac{\overline{e}^{1/2}}{S} \approx \frac{c u_*}{S} = \frac{c \kappa z}{\phi_m},
\end{equation}
where, $u_*$ and $\phi_m$ denote surface friction velocity and non-dimensional velocity gradient, respectively. $\kappa$ is the von K\'{a}rm\'{a}n constant. Based on data from the Cabauw tower in the Netherlands, \cite{nieuwstadt84} reported the proportionality constant $c$ to be approximately equal to 2.1. A similar value was also reported by \cite{basu06}.  

Since $L_H$ is proportional to $\kappa z$ in the surface layer, it can be directly compared with the so-called master length scale ($L_M$) of \citeauthor{mellor82}~\cite{mellor82}. They proposed:
\begin{equation}
    \overline{\varepsilon} = \frac{q^3}{B_1 L_M},
    \label{MY82}
\end{equation}
where, $q$ equals $\left(2 \overline{e} \right)^{1/2}$ and $B_1$ is a constant. Various forms of $L_M$ exist in the literature; however, all of them reduce to $\kappa z$ in the surface layer.

If we replace $L_M$ with $L_H$ in Eq.~\ref{MY82}, then by utilizing Eq.~\ref{EDR1}, we arrive at: 
\begin{equation}
    B_1 = \frac{q^3}{0.23 \overline{e}^{3/2}} = 12.3. 
\end{equation}
Based on various observational data, \cite{mellor82} recommended $B_1$ to be equal to 16.6. By using data from large-eddy simulations, \cite{nakanishi01} recommended $B_1 = 24.0$. Interestingly, \citeauthor{janjic02}~\cite{janjic02} heuristically derived $B_1 = 11.877992$ \citep[see also][]{foreman12}. This value of $B_1$ is currently used in the popular MYJ planetary boundary-layer scheme of the Weather Research and Forecasting (WRF) model. It is quite a coincidence that our DNS-based result turn out to be almost identical to an earlier proposition by Janji\'{c}. 

\section{Concluding Remarks}

The boundary-layer community almost always utilizes buoyancy-based energy dissipation rate parametrizations for numerical modelling studies.  Our DNS-based results suggest that shear-based parametrizations are more appropriate for regions of the stable boundary layer where $Ri_g$ does not exceed 0.2. This finding is in complete agreement with the theoretical work (supported by numerical results) of \citeauthor{hunt88}~\cite{hunt88}. They concluded: 
\begin{quote}
    ``...when the Richardson number is less than half, it is the mean shear ... (rather than the buoyancy forces) which is the dominant factor that determines the spatial velocity correlation functions and hence the length scales which determine the energy dissipation or rate of energy transfer from large to small scales.''
\end{quote}
Hunt's hypothesis was recently supported by \citeauthor{mater14}~\cite{mater14}. Through rigorous analyses of DNS and laboratory data, they found that the length scale of the overturning motions in the shear-dominated regime scale with $L_H$, whereas, in the buoyancy-dominated region, they scale with $L_b$. In addition, by utilizing observations from two well-known boundary layer field campaigns (CASES-99 and Surface Heat Budget of the Arctic Ocean \textemdash SHEBA), \citeauthor{wilson15}~\cite{wilson15} also found that $L_H$ is more correlated with the classical mixing length in comparison with the buoyancy length scale. They proposed shear-based eddy-viscosity and eddy-diffusivity parameterizations and showed promising results in an idealized simulation.    

In our future modeling studies (including large-eddy simulations), we intend to combine both the shear-based and buoyancy-based length scale parameterizations in a physically meaningful way. Simple interpolation approaches already exist in the literature \citep[e.g.,][]{grisogono08,rodier17}. An alternative approach would be to utilize a length scale proposed by \citeauthor{cheng94}~\cite{cheng94} as it seems to capture the traits of both the shear-based and buoyancy-based length scales. We are currently exploring these possibilities and others.

\section*{Data and Code Availability}
The DNS code (HERCULES) is available from: \url{https://github.com/friedenhe/HERCULES}. All the analysis codes and processed data are publicly available at \url{http://doi.org/10.5281/zenodo.3923649}. Given the sheer size of the raw DNS dataset, it is not uploaded onto any repository; however, it is available upon request from the authors.  

\begin{acknowledgments}
The first author thanks Bert Holtslag for thought-provoking discussions on this topic. The quality of the manuscript was improved by the valuable suggestions of four anonymous reviewers. We are indebted to one of the reviewers for pointing us to a possible connection of our energy dissipation rate formulation and the well-known $B_1$ constant of the MYJ planetary boundary-layer scheme. The authors acknowledge computational resources obtained from the Department of Defense Supercomputing Resource Center (DSRC) for the direct numerical simulations. The views expressed in this paper do not reflect official policy or position by the U.S Air Force or the U.S. Government.
\end{acknowledgments}

\appendix

\section*{Appendix 1: Derivation of Length Scales}

\paragraph{Integral Length Scale:}

Based on the original ideas of \citeauthor{taylor35}~\cite{taylor35}, both \cite{tennekes72}  and \cite{pope00} provided a heuristic derivation of the integral length scale. Given TKE ($\overline{e}$) and mean energy dissipation rate ($\overline{\varepsilon}$), an associated integral time scale can be approximated as $\overline{e}/\overline{\varepsilon}$. One can further assume $\sqrt{\overline{e}}$ to be the corresponding velocity scale. Thus, an integral length scale ($\mathcal{L}$) can be approximated as $\overline{e}^{3/2}/\overline{\varepsilon}$. 

In the literature, the autocorrelation function of the longitudinal velocity series is commonly used to derive an estimate of the integral length scale ($L_{11}$). The relationship between $\mathcal{L}$ and $L_{11}$ is discussed by \cite{pope00}.

\paragraph{Kolmogorov Length Scale:}

\citeauthor{pope00}~\cite{pope00} paraphrased the first similarity hypothesis of \cite{kolmogorov41a} as (the mathematical notations were changed by us for consistency): 
\begin{quote}
    ``In every turbulent flow at sufficiently high Reynolds number, the statistics of the small-scale motions ($l \ll \mathcal{L}$) have a universal form that is uniquely determined by $\nu$ and $\overline{\varepsilon}$.''  
\end{quote}
Based on $\nu$ and $\overline{\varepsilon}$, the following length scale can be formulated using dimensional analysis: $\eta \equiv \left(\frac{\nu^3}{\overline{\varepsilon}}\right)^{1/4}$. At this scale, TKE is converted into heat by the action of molecular viscosity. 

\paragraph{Ozmidov Length Scale:} \citeauthor{dougherty61}~\cite{dougherty61} and \citeauthor{ozmidov65}~\cite{ozmidov65} independently proposed this length scale. Here, we briefly summarize the derivation of \cite{ozmidov65}. Based on \cite{kolmogorov41a}, the first-order moment of the velocity increment ($\mathrm{\Delta} u$) in the vertical direction ($z$) can be written as: 
\begin{equation}
    \overline{u\left(z+\mathrm{\Delta} z\right) - u(z)} = \overline{\mathrm{\Delta} u} = \mathrm{\Delta} \overline{u} \approx \overline{\varepsilon}^{1/3} \mathrm{\Delta} z^{1/3},
\end{equation}
where the overlines denote ensemble averaging. Using this equation, the vertical gradient of longitudinal velocity component can be approximated as: 
\begin{equation}
    \frac{\partial \overline{u}}{\partial z} \approx \frac{\mathrm{\Delta} \overline{u}}{\mathrm{\Delta} z} \approx \overline{\varepsilon}^{1/3} \mathrm{\Delta} z^{-2/3}.
\end{equation}
Similar equation can be written for the vertical gradient of the lateral velocity component ($\frac{\partial \overline{v}}{\partial z}$). Thus, the magnitude of wind shear ($S$) can be written as: 
\begin{equation}
    S \approx \overline{\varepsilon}^{1/3} \mathrm{\Delta} z^{-2/3}. 
\end{equation}
By definition, $Ri_g = N^2/S^2$. Thus,
\begin{equation}
    Ri_g \approx \frac{N^2}{\overline{\varepsilon}^{2/3} \mathrm{\mathrm{\Delta}} z^{-4/3}}.
    \label{Rig_OZ}
\end{equation}
\citeauthor{ozmidov65}~\cite{ozmidov65} assumed that for a certain critical $Ri_g$ (which is assumed to be an unknown constant), $\mathrm{\Delta} z$ becomes the representative outer length scale ($L_{OZ}$). Thus, Eq.~\ref{Rig_OZ} can be rewritten as: 
\begin{equation}
    L_{OZ} \equiv \left(\frac{\overline{\varepsilon}}{N^3}\right)^{1/2}.
\end{equation}
The unknown proportionality constant is a function of the critical $Ri_g$ and is assumed to be on the order of one.

\paragraph{Corrsin Length Scale:}

The derivation of \citeauthor{corrsin58}~\cite{corrsin58} leverages on a characteristic spectral time scale,  $T_s(\kappa)$, which is representative of the inertial range. Based on dimensional argument, \citeauthor{onsager49}~\cite{onsager49} proposed:   
\begin{equation}
    T_s(\kappa) \equiv \frac{1}{\sqrt{\kappa^2E(\kappa)}}.
\end{equation}
 In order to guarantee local isotropy in the inertial-range, \citeauthor{corrsin58}~\cite{corrsin58} hypothesized that $T_s(\kappa)$ must be much smaller than the time scale associated with mean shear ($S$). In other words,
\begin{equation}
    \frac{1}{\sqrt{\kappa^2E(\kappa)}} \ll \frac{1}{S}.
\end{equation}
Using the -5/3 law of \cite{kolmogorov41a} and \cite{obukhov41a,obukhov41b}, this equation can be rewritten as: 
\begin{equation}
    \frac{1}{\sqrt{\kappa^{4/3} \overline{\varepsilon}^{2/3}}} \ll \frac{1}{S}.
    \label{LC1}
\end{equation}
If we assume that for a specific wavenumber $\kappa = 1/L_C$, the equality holds in  Eq.~\ref{LC1}, then we get: 
\begin{equation}
    L_C^{2/3} = \frac{\overline{\varepsilon}^{1/3}}{S}.
\end{equation}
From this equation, we can estimate $L_C$ as defined earlier in Eq.~\ref{LC}.

\paragraph{Buoyancy Length Scale:}

The following heuristic derivation is based on \cite{brost78} and \cite{wyngaard10}. In an order-of-magnitude analysis, the inertia term  of the Navier\textendash Stokes equations, can be written as: 
\begin{equation}
    \frac{\partial u_i}{\partial t} \sim \frac{U_s}{T_s} \sim \frac{U_s}{L_s/U_s} \sim \frac{U_s^2}{L_s}, 
    \label{LB1}
\end{equation}
where $L_s$, $T_s$, and $U_s$ represent certain length, time, and velocity scales, respectively. In a similar manner, the buoyancy term can be approximated as: 
\begin{equation}
    \left(\frac{g}{\Theta_0}\right)\left(\theta'\right) \sim \left(\frac{g}{\Theta_0}\right)\left(\frac{\partial \overline{\theta}}{\partial z}\right) \left(L_s\right) \sim N^2 L_s,
    \label{LB2}
\end{equation}
where $\Theta_0$ and $\theta'$ denote a reference temperature and temperature fluctuations, respectively. Equating the inertia and the buoyancy terms, we get: 
\begin{equation}
    L_s^2 = \frac{U_s^2}{N^2}.
\end{equation}
For stably stratified flows, either $\overline{e}^{1/2}$ or $\sigma_w$ can be used as an appropriate velocity scale. Accordingly, the length scale ($L_s$) can be approximated as $\frac{\overline{e}^{1/2}}{N}$ or $\frac{\sigma_w}{N}$. In the literature, this length scale is commonly known as the buoyancy length scale ($L_b$). 

\paragraph{Hunt Length Scale:}

\citeauthor{hunt88}~\cite{hunt88} hypothesized that in stratified shear flows, $\overline{\varepsilon}$ is controlled by mean shear ($S$) and $\sigma_w$. From dimensional analysis, it follows that: 
\begin{equation}
    \overline{\varepsilon} \equiv \sigma_w^2 S.
\end{equation}
The associated length scale, $L_H$, is assumed to be on the order of $\sigma_w/S$. 

\section*{Appendix 2: Energy Dissipation Rate Formulation by Weinstock}

The starting point of Weinstock's derivation was the -5/3 law of \cite{kolmogorov41a} and \cite{obukhov41a,obukhov41b}. He integrated this equation in the wavenumber space and set the upper integration limit to infinity. The lower integration limit was fixed at the buoyancy wavenumber ($\kappa_b$). Furthermore, he assumed that the eddies are isotropic for wavenumbers larger than $\kappa_b$ (i.e., in the inertial and viscous ranges). His derivation can be summarized as: \begin{equation}
\begin{split}
    \frac{3}{2} \sigma_w^2 & = \int_{\kappa_b}^{\kappa_2} \alpha \overline{\varepsilon}^{2/3} \kappa^{-5/3} d\kappa \\
    & = \alpha \overline{\varepsilon}^{2/3} \int_{\kappa_b}^{\kappa_2} \kappa^{-5/3} d\kappa \\
    & = \frac{3 \alpha}{2} \overline{\varepsilon}^{2/3} \left(\kappa_b^{-2/3} - \kappa_2^{-2/3} \right) \\
    & \approx \frac{3 \alpha}{2} \overline{\varepsilon}^{2/3} \kappa_b^{-2/3}.
\end{split}
\label{W81x}
\end{equation}
\citeauthor{weinstock81}~\cite{weinstock81} assumed that $\kappa_b$ can be parametrized by $\frac{N}{\sigma_w}$ (basically, the inverse of the buoyancy length scale $L_b$). By plugging this parametrization into Eq.~\ref{W81x} and simplifying, we get: 
\begin{equation}
\begin{split}
\overline{\varepsilon} & \approx \sigma_w^3 \kappa_b \\
& \approx \sigma_w^2 N.
\end{split}
\end{equation}

\section*{Appendix 3: Normalization of Variables in Direct Numerical Simulations}
\label{appendix3}

In DNS, the relevant variables are normalized as follows: 
\begin{subequations}
\begin{equation}
z_n = \frac{z}{h}, 
\end{equation}
\begin{equation}
u_n = \frac{u}{U_b}, 
\end{equation}
\begin{equation}
v_n = \frac{v}{U_b}, 
\end{equation}
\begin{equation}
w_n = \frac{w}{U_b}, 
\end{equation}
\begin{equation}
\theta_n = \frac{\theta-\Theta_{top}}{\Theta_{top}-\Theta_{bot}}. 
\end{equation}
\end{subequations}
After differentiation, we get: 
\begin{subequations}
\begin{equation}
\frac{\partial u}{\partial z} =  \frac{\partial u}{\partial z_n} \frac{\partial z_n}{\partial z} = \frac{\partial u}{\partial u_n} \frac{\partial u_n}{\partial z_n}\frac{\partial z_n}{\partial z} = \frac{U_b}{h} \frac{\partial u_n}{\partial z_n}, 
\end{equation}
\begin{equation}
\frac{\partial v}{\partial z} =  \frac{\partial v}{\partial z_n} \frac{\partial z_n}{\partial z} = \frac{\partial v}{\partial v_n} \frac{\partial v_n}{\partial z_n}\frac{\partial z_n}{\partial z} = \frac{U_b}{h} \frac{\partial v_n}{\partial z_n}, 
\end{equation}
\begin{equation}
S =  \sqrt{\left(\frac{\partial \overline{u}}{\partial z}\right)^2 + \left(\frac{\partial \overline{v}}{\partial z} \right)^2} = \frac{U_b}{h} S_n,
\end{equation}
\begin{equation}
\frac{\partial \theta}{\partial z} =  \frac{\partial \theta}{\partial z_n} \frac{\partial z_n}{\partial z} = \frac{\partial \theta}{\partial \theta_n}
\frac{\partial \theta_n}{\partial z_n}\frac{\partial z_n}{\partial z} = \left(\frac{\Theta_{top}-\Theta_{bot}}{h}\right) \frac{\partial \theta_n}{\partial z_n}. 
\end{equation}
\end{subequations}

\begin{widetext}
The gradient Richardson number can be expanded as: 
\begin{equation}
    Ri_g = \frac{N^2}{S^2} = \frac{\left(\frac{g}{\Theta_0}\right)\left(\frac{\partial \overline{\theta}}{\partial z}\right)}{S^2}
    = \left(\frac{g}{\Theta_{top}}\right) \left(\frac{\Theta_{top}-\Theta_{bot}}{h}\right)\left(\frac{h}{U_b} \right)^2 \frac{\left(\frac{\partial \overline{\theta}_n}{\partial z_n} \right)}{S_n^2}.
\end{equation}
\end{widetext}
Using the definition of $Ri_b$ (see Sect.~2), we rewrite $Ri_g$ as follows:
\begin{equation}
    Ri_g = Ri_b \frac{\left(\frac{\partial \overline{\theta}_n}{\partial z_n} \right)}{S_n^2}.
\end{equation}
Similarly, $N^2$ can be written as: 
\begin{equation}
    N^2 = Ri_b \left(\frac{U_b^2}{h^2}\right) \left(\frac{\partial\overline{\theta}_n}{\partial z_n}\right).
    \label{EqN2}
\end{equation}
The velocity variances and TKE can be normalized as: 
\begin{subequations}
\begin{equation}
\sigma_{u_n}^2 = \frac{\sigma_u^2}{U_b^2},
\end{equation}
\begin{equation}
\sigma_{v_n}^2 = \frac{\sigma_v^2}{U_b^2},
\end{equation}
\begin{equation}
\sigma_{w_n}^2 = \frac{\sigma_w^2}{U_b^2},
\label{EqSigw}
\end{equation}
\begin{equation}
\overline{e}_n = \frac{\overline{e}}{U_b^2}. 
\label{Eqe}
\end{equation}
\end{subequations}
Following the above normalization approach, we can also derive the following relationship for the energy dissipation rate: 
\begin{equation}
    \overline{\varepsilon} = \nu \left(\frac{U_b}{h}\right)^2 \overline{\varepsilon}_n. 
    \label{EqEDR}
\end{equation}
In order to expand $\overline{\varepsilon} = \overline{e} N$, we use Eq.~\ref{EqN2}, Eq.~\ref{Eqe}, and Eq.~\ref{EqEDR} as follows: 
\begin{equation}
    \nu \left(\frac{U_b}{h}\right)^2 \overline{\varepsilon}_n = U_b^2 \overline{e}_n
    Ri_b^{1/2} \left(\frac{U_b}{h}\right) \left(\frac{\partial\overline{\theta}_n}{\partial z_n}\right)^{1/2}.
\end{equation}
This equation can be simplified to: 
\begin{equation}
\overline{\varepsilon}_n = Re_b Ri_b^{1/2} \overline{e}_n \left(\frac{\partial\overline{\theta}_n}{\partial z_n}\right)^{1/2}.
\end{equation}
In a similar manner, $\overline{\varepsilon} = \overline{e} S$ can be re-written as: \begin{equation}
\overline{\varepsilon}_n = Re_b \overline{e}_n S_n.    
\end{equation}

\section*{Appendix 4: Supplementary Analyses of Simulated Data}

\begin{figure*}
\centering
  \includegraphics[width=1.5in,height=1.5in]{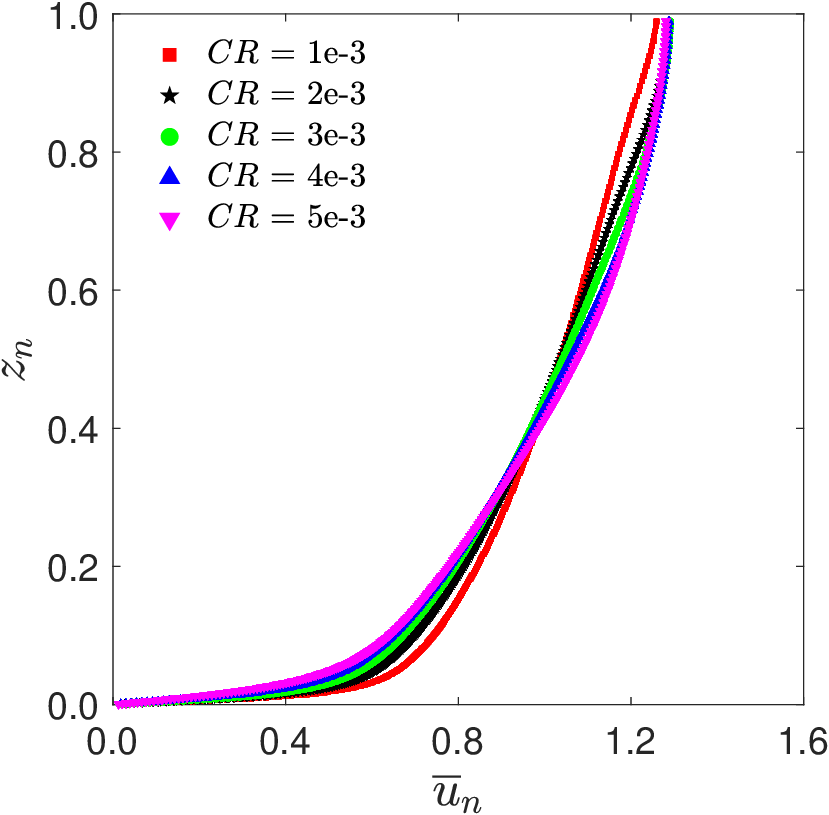}
  \hspace{0.3in}
  \includegraphics[width=1.5in,height=1.5in]{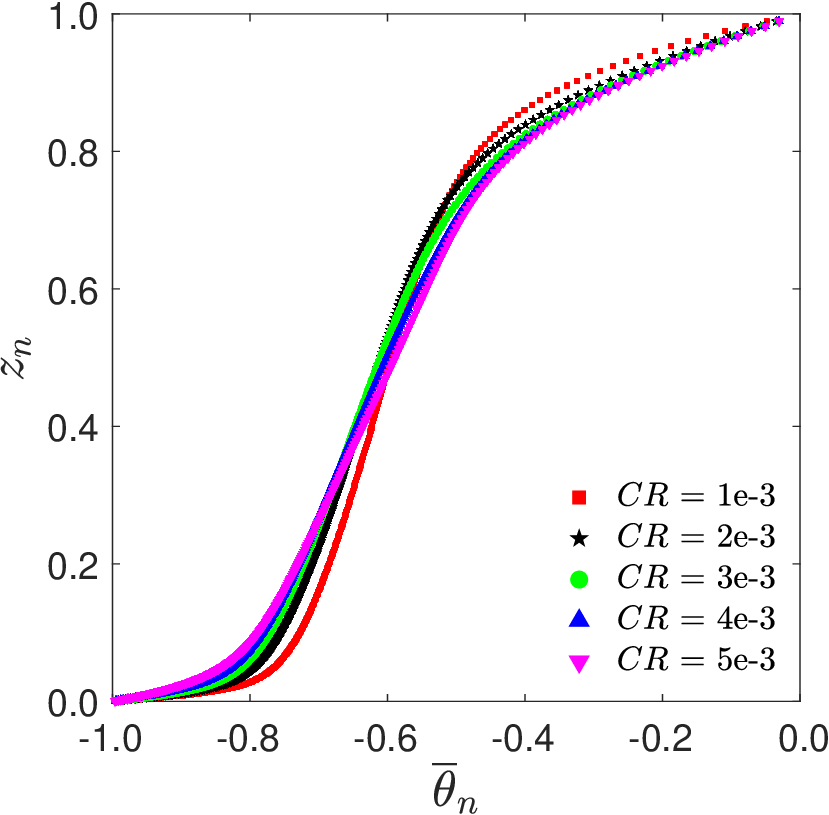}
  \hspace{0.3in}
  \includegraphics[width=1.5in,height=1.5in]{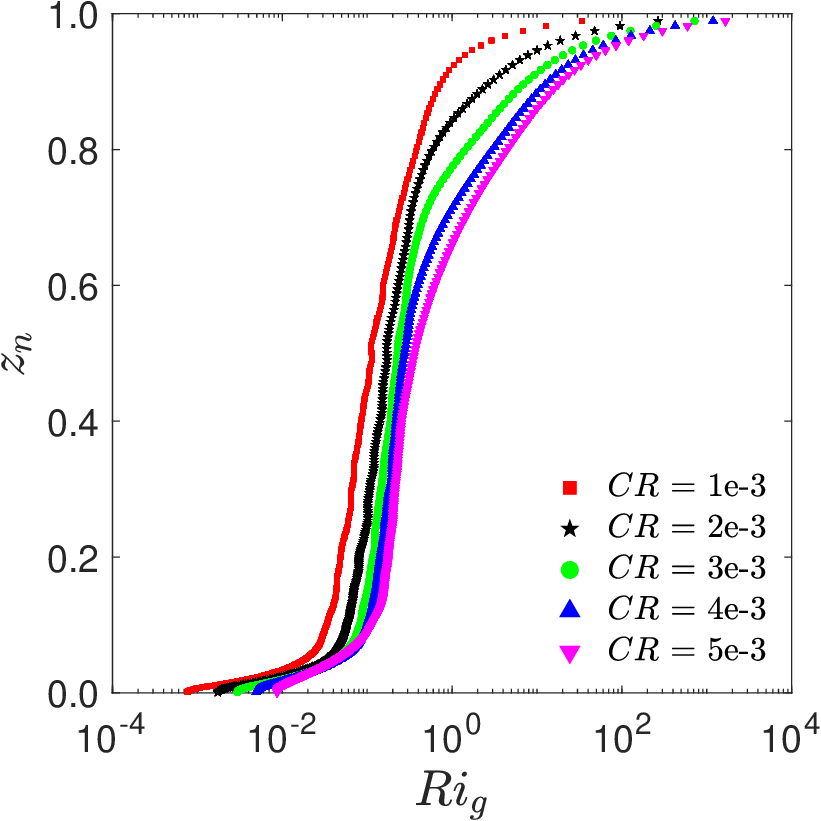}\\
  \includegraphics[width=1.5in,height=1.5in]{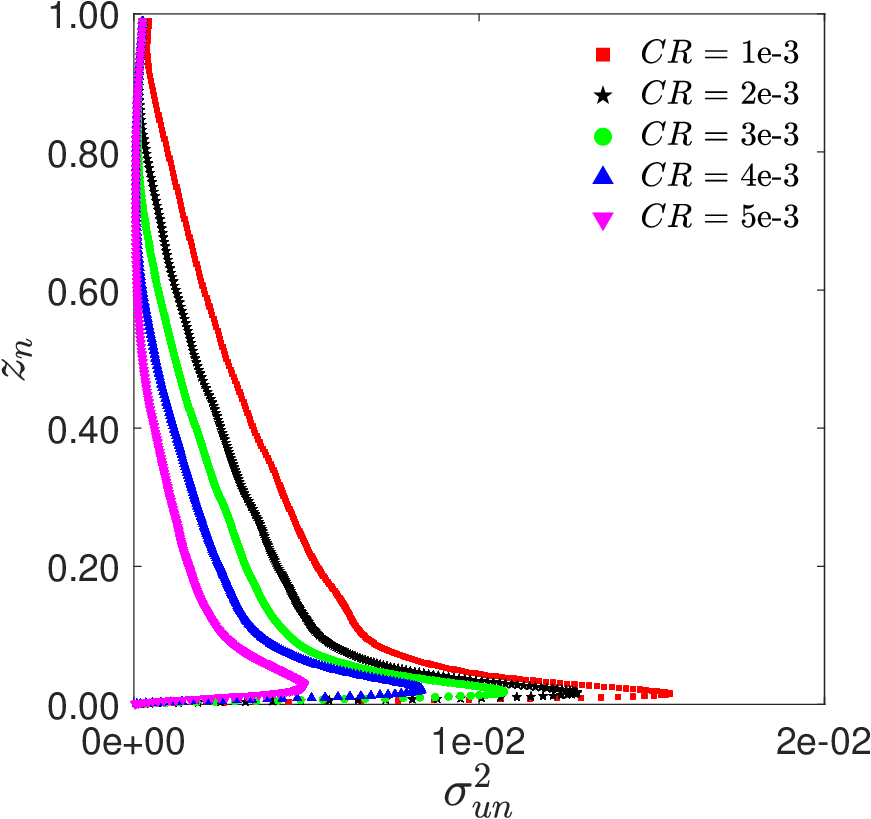}
  \hspace{0.3in}
  \includegraphics[width=1.5in,height=1.5in]{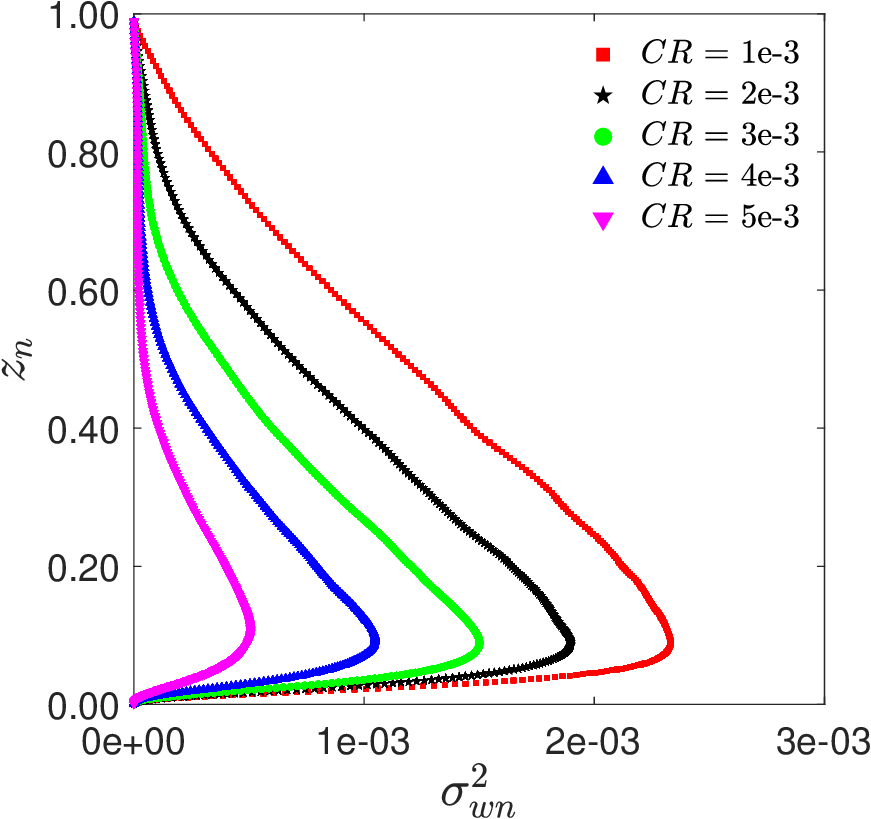}
  \hspace{0.3in}
  \includegraphics[width=1.5in,height=1.5in]{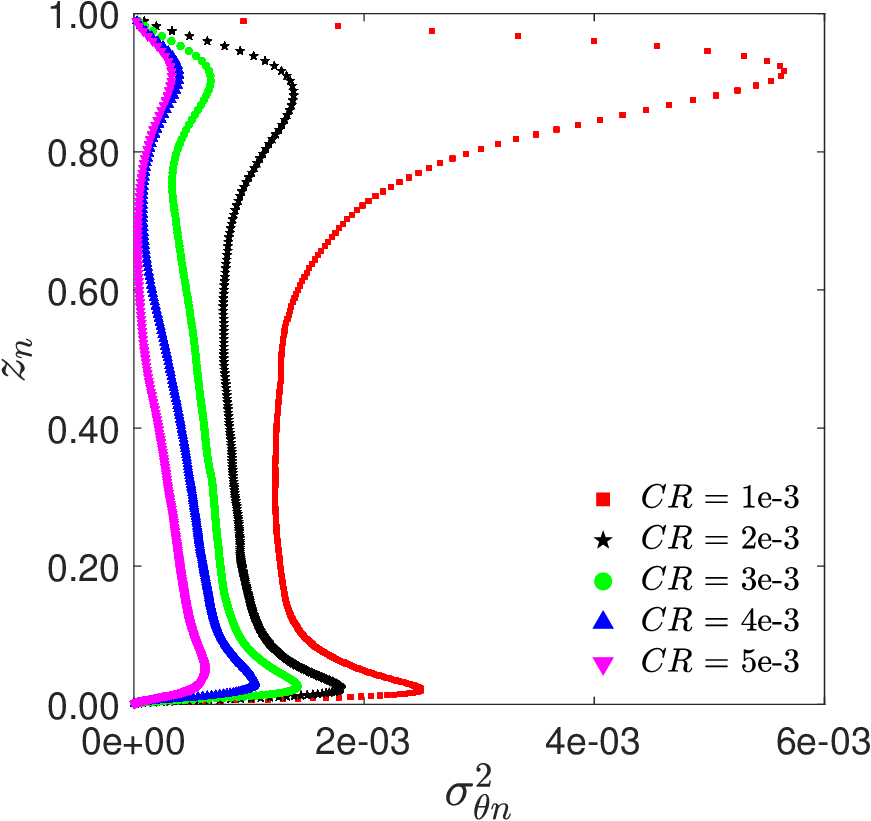}\\
  \includegraphics[width=1.5in,height=1.5in]{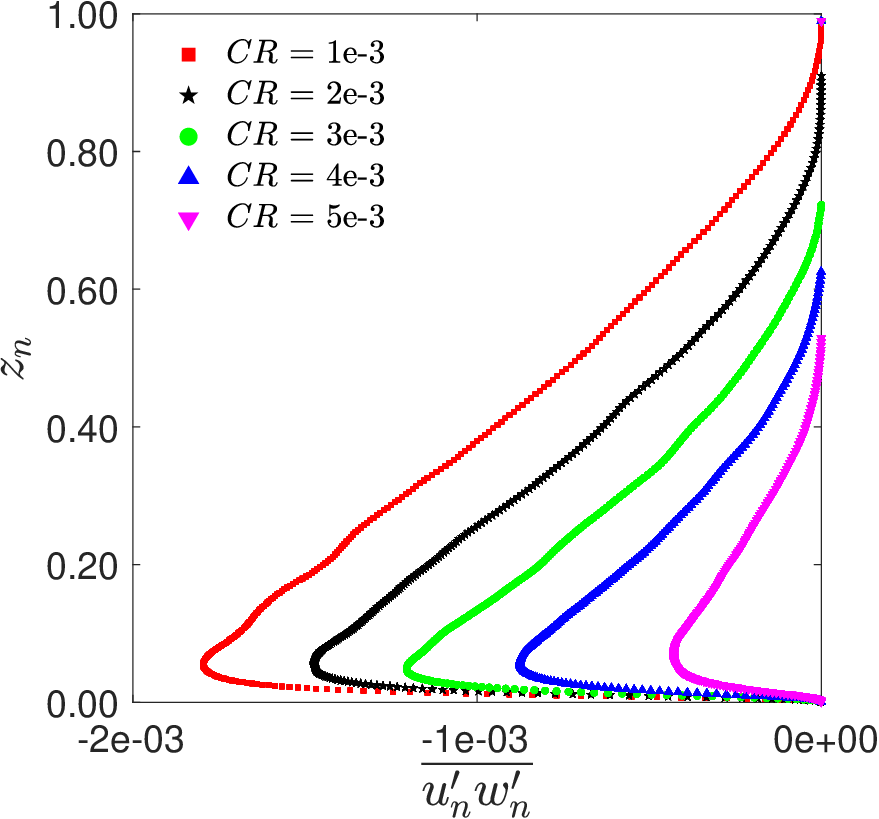}
  \hspace{0.3in}
  \includegraphics[width=1.5in,height=1.5in]{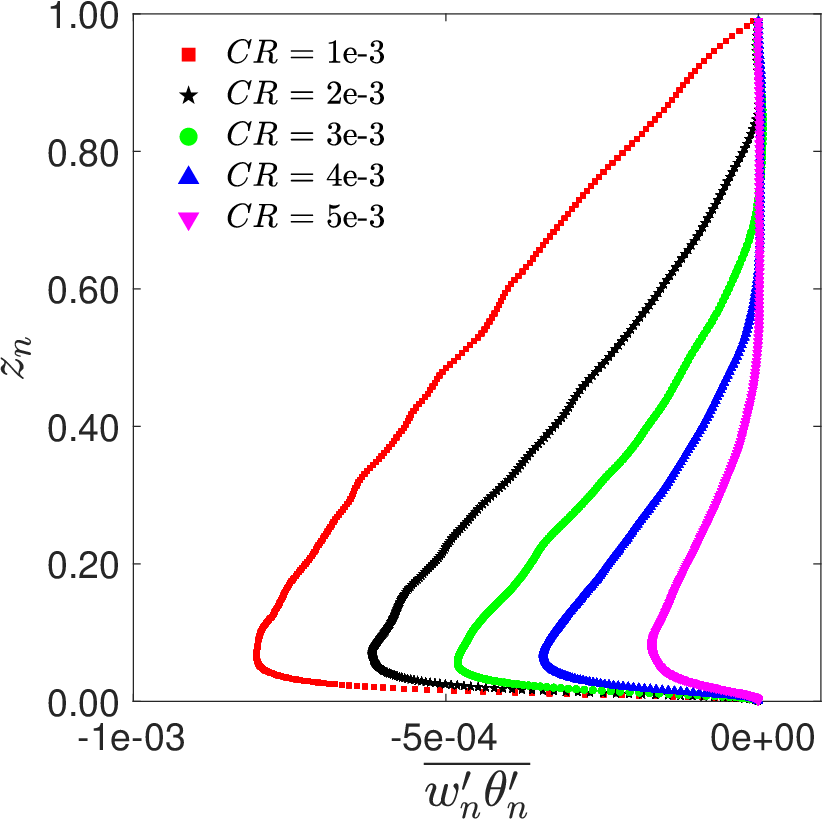}
  \hspace{0.3in}
  \includegraphics[width=1.5in,height=1.5in]{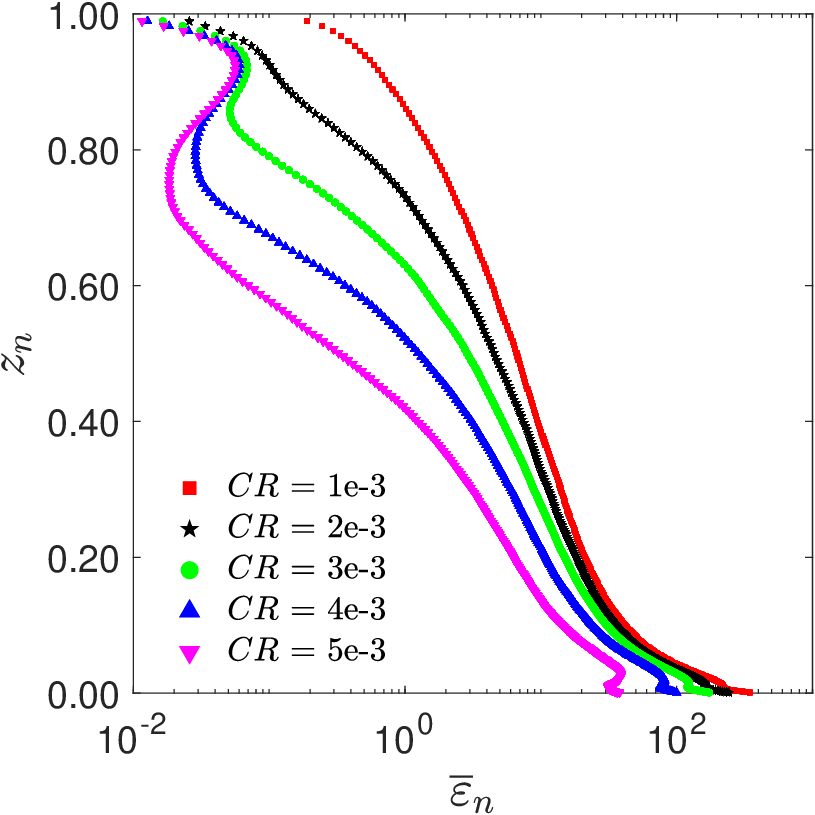}
\caption{Vertical profiles of normalized longitudinal velocity (top-left panel), potential temperature (top-center panel), gradient Richardson number (top-right panel), longitudinal velocity variance (middle-left panel), vertical velocity variance (middle-center panel), potential temperature variance (middle-right panel), $u$-component of momentum flux (bottom-left panel), sensible heat flux (bottom-center panel), and energy dissipation rate (bottom-right panel). Simulated data from five different DNS runs are represented by different coloured symbols in these plots. In the legends, $CR$ represents normalized cooling rates. All the profiles correspond to $T_n = 100$}
\label{fig5}       
\end{figure*}

In Fig.~\ref{fig5}, vertical profiles of several key variables are plotted. All the profiles correspond to $T_n = 100$. Clearly, the variances and fluxes decrease with increasing cooling rate. It is also evident that stability monotonically increases with height. As a result, turbulence in the upper part of the domain becomes quasi-laminar (especially for the runs with higher cooling rates). For this reason, we did not consider data from $z/h > 0.5$ region for the computations of various length scales.

For continuously turbulent SBLs, it has been frequently observed that $Ri_g$ stays below 0.2 within the SBL \citep[e.g.,][]{garratt82,nieuwstadt84,basu06}. Above the SBL, in the free atmosphere, $Ri_g$ becomes much larger. Similar behaviour is noticeable in Fig.~\ref{fig5} (top-right panel).

The vertical profiles of dissipation rates are shown in the bottom-right panel of Fig.~\ref{fig5}. As expected, the dissipation rates decrease with increasing height. For $z/h < 0.1$, due to the viscous effects, the values of the dissipation rates are very high. Thus, for analyses of the length scales, we disregarded data from this region. 

\bibliography{EDR}

\end{document}